\documentclass[twocolumn,english,nofootinbib]{revtex4}
\usepackage[latin9]{inputenc}
\usepackage{amssymb}

\makeatletter

\usepackage{babel}
\makeatother

\begin{document}

\title{Backreaction of superhorizon perturbations in scalar field cosmologies}

\author{Naresh Kumar}

\email{nk236@cornell.edu}

\author{$\acute{\mathrm{E}}$anna $\acute{\mathrm{E}}$. Flanagan}

\email{eef3@cornell.edu}

\affiliation{Laboratory of Elementary Particle Physics, Cornell University, Ithaca,
New York 14853, USA.}

\begin{abstract}
~

~~ It has been suggested that the acceleration of the Universe may
be due to the backreaction of perturbations to the Friedmann-Robertson-Walker
background. For a Universe dominated by cold dark matter, it is known
that the backreaction of superhorizon perturbations can not drive
acceleration. We extend this result to models with cold dark matter
together with a scalar field. We show that the scalar field can drive
acceleration only via the standard mechanism of a constant or nearly
constant piece of its potential (i.e., a cosmological constant); there
is no separate mechanism involving superhorizon backreaction. This
rules out some models which have been proposed in the literature.

~
\end{abstract}
\maketitle

\section{Introduction and Summary}

\ ~The nature of dark energy is one of the most important outstanding
problems in cosmology. A simple explanation is achieved by inserting
a cosmological constant in Einstein's field equation. However, there
are well known naturalness and fine tuning problems associated with
a cosmological constant {[}1]. The simplest models offering a solution
to the fine tuning problem are quintessence models {[}2]. Cosmic acceleration
can also in principle be explained by modifying general relativity
at large distance scales. Examples include f(R) theories {[}3, 4,
5, 6, 7] and the Dvali-Gabadadze-Porrati model {[}8, 9, 10, 11, 12,
13, 14]. For a description of more models see the review {[}15] and
references within. The naturalness problem persists in most of these
dynamical models.

It has also been suggested that the acceleration of the Universe can
be explained by a purely general relativistic effect involving no
new physics, the backreaction of perturbations {[}16, 17, 18, 19,
20]. See Refs. {[}17, 18, 21, 22, 23, 24] for a review of the backreaction
idea. By taking a spatial average of Einstein's equations in a particular
gauge, one can obtain Friedmann equations for an effective spatially
averaged scale factor $a\left(t\right)$ with extra driving terms
coming from backreaction {[}25, 26, 27, 28, 29]. These extra driving
terms can in principle drive an acceleration. Although the conventional
viewpoint is that the effect of backreaction is small, some have argued
that it can be large enough to account for cosmic acceleration.

A problem with this theoretical approach is that the spatially averaged
scale factor is not related in any simple way to quantities we observe,
which average over our past light cone. See Refs. {[}22, 30] for a
discussion of this issue.

There are two variants of the backreaction explanation. The first
is that cosmic acceleration is caused by the backreaction of $\mathit{superhorizon}$
perturbations. In particular Kolb et al {[}16, 17] looked at inflation-generated
perturbations to a Friedmann-Robertson-Walker (FRW) universe, and
claimed that at second order one could obtain a negative deceleration
parameter. This claim was disproved in Refs. {[}31 - 33] %
\footnote{We also point to Ref. {[}34] for other arguments against cosmic acceleration
caused by backreaction%
}.

The second variant of the backreaction explanation is that the backreaction
of $\mathit{subhorizon}$ perturbations can explain cosmic acceleration
{[}18, 35, 36, 37, 38]. This seems unlikely but the issue has not
yet been settled. We will not discuss the subhorizon backreaction
issue here. 

In this paper we focus on the backreaction of superhorizon perturbations
in the present day Universe %
\footnote{We note that there is also a considerable literature on superhorizon
backreaction during the inflationary era, which does have a local
physical effect in two scalar field inflation models {[}39, 40, 41,
42, 43, 44, 45, 46, 47, 48 ,49].%
}. In particular we show that for a Universe with cold dark matter
and a scalar field (as in standard quintessence models), achieving
a value $q_{0}\simeq-0.5$ of the deceleration parameter requires
a non-zero potential. Our method of analysis is as follows {[}31,
32]. We compute luminosity distance as a function of redshift using
Taylor series expansions, in an arbitrary Universe containing cold
dark matter and a scalar field. By angle averaging we then infer the
observed value of the deceleration parameter $q_{0}$. Our result
shows that if such a Universe is accelerating, that acceleration must
be primarily driven by the standard mechanism of a cosmological constant
term in the scalar field's potential. If the potential is absent,
the backreaction of superhorizon perturbation of the scalar field
can not drive acceleration. In particular, second order perturbations
are not sufficient to explain cosmic acceleration. 

Our analysis was motivated in part by a claim by Martineau and Brandenberger
{[}50] that the acceleration could be caused by the backreaction of
superhorizon perturbations of a scalar field. These authors consider
a model in which a single scalar field both drives inflation and is
also present today. Modeling the scalar field perturbations using
an effective energy momentum tensor, they argue that the effect of
those perturbations can be of the right magnitude and character to
cause acceleration. Our result shows that this model cannot be correct.
We discuss further in Sec. III below a possible reason for our differing
results.

\section{Computation of Deceleration Parameter}

~~~We start by describing our theoretical framework and assumptions,
which are a slightly modified version of those used in Ref. {[}31].
We consider the Universe in the matter dominated era, described by
general relativity coupled to a pressureless fluid describing cold
dark matter (we neglect baryons and radiation), together with a light
scalar field. Our starting point is the assumption that backreaction
is dominated by the effect of superhorizon perturbations. If this
is true, then backreaction should also be present in a hypothetical,
gedanken Universe in which all the perturbation modes which are subhorizon
today are set to zero at early times. Generic solutions to the field
equations for this gedanken Universe can be described using local
Taylor series expansions rather than via perturbations of Friedmann-Robertson-Walker
models, since all the fields vary on length scales or time scales
of order the Hubble time or larger. This greatly simplifies the analysis.

The three equations which describe the dynamics of the gedanken universe
are the following:

\[
G_{\alpha\beta}=8\pi\biggl[\left(\rho+p\right)u_{\alpha}u_{\beta}+pg_{\alpha\beta}+\nabla_{\alpha}\phi\nabla_{\beta}\phi\]

\begin{equation}
-\frac{1}{2}g_{\alpha\beta}\left(\nabla\phi\right)^{2}-V\left(\phi\right)g_{\alpha\beta}\biggr],\label{eq:1}\end{equation}

\begin{equation}
\square\phi-V^{'}\left(\phi\right)=0,\label{eq:2}\end{equation}
and

\begin{equation}
\nabla_{\alpha}\left[\left(\rho+p\right)u^{\alpha}u^{\beta}+pg^{\alpha\beta}\right]=0.\label{eq:3}\end{equation}
Here $\rho$, $p$ and $u_{\alpha}$ are the density, pressure and
four velocity of matter, $\phi$ is the scalar field and $V\left(\phi\right)$
is its potential. Later we will specialize to cold dark matter for
which $p=0.$

Next we define the specific deceleration parameter $q_{0}$ that we
use. As discussed in Ref. {[}33], there are several different possible
definitions for non-FRW cosmological models. The definition we choose
matches closely with how $q_{0}$ is actually measured. 

Let us start by fixing a comoving observer at point $O$ in spacetime.
We label null geodesics on $O's$ past null cone in terms of the spherical
polar angles ($\theta$,$\phi$) of a local Lorentz frame at $O$
that is comoving with the cosmological fluid. We parameterize each
null geodesic in terms of an affine parameter $\lambda$ and corresponding
4-momentum $\vec{k}=d/d\lambda$. For a given source $S$ on such
a null geodesic at affine parameter $\lambda$, we define the redshift
as

\begin{equation}
1+z\left(\theta\textrm{,}\phi\textrm{,}\lambda\right)=\frac{\vec{k}.\vec{u}|_{S}}{\vec{k}.\vec{u}|_{O}}.\label{eq:4}\end{equation}
The luminosity distance $\mathcal{{\normalcolor D}_{L}}\left(\theta\textrm{,}\phi\textrm{,}\lambda\right)$
of the source $S$ is defined in the usual way in terms of the luminosity
$\left(dE/dt\right)_{S}$ of an assumed comoving isotropic source
at S and the energy per unit area per unit time $\left(dE/dtdA\right)_{O}$
measured at $O$:

\begin{equation}
\left(\frac{dE}{dtdA}\right)_{O}=\frac{1}{4\pi\mathcal{D_{L}}^{2}}\left(\frac{dE}{dt}\right)_{S}.\label{eq:5}\end{equation}
Assuming that the wavelength of the radiation from $S$ is much smaller
than the radius of curvature of spacetime, we can use geometric optics
to compute the observed energy flux in Eq. (5) and thus the luminosity
distance $\mathcal{D_{L}}$; see, for example Ref. {[}51]. Finally,
we can eliminate the affine parameter $\lambda$ between Eqs. (4)
and (5) and compute the luminosity distance as a function of spherical
coordinates and redshift to obtain $\mathcal{D}_{\mathcal{L}}=\mathcal{D_{L}}\left(\theta\textrm{,}\phi\textrm{,}z\right)$. 

Next, to define the deceleration parameter $q_{0}$, we expand the
luminosity distance in powers of redshift. The result is

\begin{equation}
\mathcal{D_{L}}\left(\theta\textrm{,}\phi\textrm{,}z\right)=A\left(\theta,\phi\right)z+B\left(\theta,\phi\right)z^{2}+O\left(z^{3}\right),\label{eq:6}\end{equation}
where $A\left(\theta,\phi\right)$ and \textbf{$B\left(\theta,\phi\right)$}
are functions that only have angular dependences. We then define the
Hubble parameter $H_{0}$ and the deceleration parameter $q_{0}$
in terms of angular averages of the above functions. The standard
FRW relation is \begin{equation}
\mathcal{D_{L}}\left(\theta\textrm{,}\phi\textrm{,}z\right)=H_{0}^{-1}z+H_{0}^{-1}\left(1-q_{0}\right)\frac{z^{2}}{2}+O\left(z^{3}\right).\label{eq:7}\end{equation}
Comparing the expansions (6) and (7) motivates the following definitions
of $H_{0}$ and $q_{0}$:

\begin{equation}
H_{0}\equiv\left\langle A^{-1}\right\rangle ,\qquad q_{0}\equiv1-2H_{0}^{-2}\left\langle A^{-3}B\right\rangle ,\label{eq:8}\end{equation}
where $\left\langle ...\right\rangle $ denotes an average over the
angles $\theta$ and $\phi$. Note that there is some ambiguity in
these definitions. For example one could take $q_{0}=1-2\left\langle A^{-1}B\right\rangle $
instead. We choose the form (8) for computational convenience, and
we will argue below that the differences are unimportant.

We next explicitly evaluate the expressions (8) for $H_{0}$ and $q_{0}$.
We consider generic solutions to the equations (1) - (3), described
by local Taylor series expansions {[}52] about the observer $O$.
The expressions for the functions $A$ and $B$ were computed in Ref.
{[}31], and are

\begin{equation}
A\left(\theta,\phi\right)=\frac{1}{\left(\nabla^{\alpha}u^{\beta}\right)k_{\alpha}k_{\beta}},\label{eq:9}\end{equation}

\begin{equation}
B\left(\theta,\phi\right)=\frac{2}{\left(\nabla^{\alpha}u^{\beta}\right)k_{\alpha}k_{\beta}}+\frac{\left(\nabla^{\alpha}\nabla^{\beta}u^{\gamma}\right)k_{\alpha}k_{\beta}k_{\gamma}}{2\left[\left(\nabla^{\alpha}u^{\beta}\right)k_{\alpha}k_{\beta}\right]^{3}},\label{eq:10}\end{equation}
where all quantities on the right hand sides are evaluated at $O.$
By inserting the expression for $A\left(\theta,\phi\right)$ into
the definition (8) of $H_{0}$ and evaluating the angular average,
we obtain

\begin{equation}
H_{0}=\frac{1}{3}\Theta,\label{eq:11}\end{equation}
where

\begin{equation}
\Theta=\nabla_{\alpha}u^{\alpha}\label{eq:12}\end{equation}
is the expansion of the cosmological fluid. This is the same result
as was obtained in Ref. {[}31].

Next, we insert the expressions (9) and (10) for $A\left(\theta,\phi\right)$
and $B\left(\theta,\phi\right)$ into (8) to obtain

\[
\frac{1}{2}H_{0}^{2}\left(1-q_{0}\right)=2\left\langle \left[\left(\nabla^{\alpha}u^{\beta}\right)k_{\alpha}k_{\beta}\right]^{2}\right\rangle \]

\begin{equation}
+\frac{1}{2}\left\langle \left(\nabla^{\alpha}\nabla^{\beta}u^{\gamma}\right)k_{\alpha}k_{\beta}k_{\gamma}\right\rangle .\label{eq:13}\end{equation}
We now evaluate these angular averages using the same techniques as
in Ref. {[}31]. The only difference from the computation of Ref. {[}31]
arises when we eliminate a factor of the Ricci tensor $R_{\alpha\beta}$
using the field equations. Here that elimination generates extra terms
involving the scalar field, from the equation of motion (1). The final
result is

\[
q_{0}=\frac{4\pi}{3H_{0}^{2}}\left.\left[\rho+3p-V\left(\phi\right)+2\nabla_{\alpha}\phi\nabla_{\beta}\phi u^{\alpha}u^{\beta}\right]\right|_{O}\]

\begin{equation}
+\frac{1}{3H_{0}^{2}}\left.\left(a_{\alpha}a^{\alpha}+\frac{7}{5}\sigma_{\alpha\beta}\sigma^{\alpha\beta}-w_{\alpha\beta}w^{\alpha\beta}-2\nabla_{\alpha}a^{\alpha}\right)\right|_{O}.\label{eq:14}\end{equation}
Here $\sigma^{\alpha\beta}$, $w^{\alpha\beta}$ and $a^{\alpha}$
are the shear, vorticity and 4-acceleration of the fluid, defined
by

\begin{equation}
\nabla_{\alpha}u_{\beta}=\frac{1}{3}\Theta g_{\alpha\beta}+\sigma_{\alpha\beta}+w_{\alpha\beta}-a_{\beta}u_{\alpha},\label{eq:15}\end{equation}

with

\begin{equation}
\sigma_{\alpha\beta}=\sigma_{\beta\alpha}\,\,\textrm{and\,}\, w_{\alpha\beta}=-w_{\beta\alpha}.\label{eq:16}\end{equation}

\section{Discussion}

We now specialize our result (14) to a pressureless fluid. The 4-acceleration
then vanishes and we have

\[
q_{0}=\frac{4\pi G}{3H_{0}^{2}}\left.\left[\rho-V\left(\phi\right)+2\nabla_{\alpha}\phi\nabla_{\beta}\phi u^{\alpha}u^{\beta}\right]\right|{}_{O}\]

\begin{equation}
+\frac{1}{3H_{0}^{2}}\left.\left(\frac{7}{5}\sigma_{\alpha\beta}\sigma^{\alpha\beta}-w_{\alpha\beta}w^{\alpha\beta}\right)\right|{}_{O}.\label{eq:17}\end{equation}
We now argue that the only way to achieve $q_{0}\thickapprox-0.5,$
as required by observations, is to have $V\left(\phi_{0}\right)$
be large and positive, where $\phi_{0}$ is the value of $\phi$ evaluated
at the observer.

We can estimate the terms $\sigma_{\alpha\beta}\sigma^{\alpha\beta}$
and $w_{\alpha\beta}w^{\alpha\beta}$ in Eq. (17) to be $\thicksim\left(\delta v\right)^{2}/l^{2}$,
where $\delta v$ is the typical scale of peculiar velocity perturbations,
and $l$ is the scale over which the velocity varies. Since we have
assumed that subhorizon perturbations are absent we have $l\gtrsim H_{0}^{-1}$.
This implies that the contributions from these terms are of order
$\delta q_{0}\sim\left(\delta v\right)^{2}\simeq10^{-4}$. Since the
measured value of $q_{0}$ is $q_{0}\thicksim-0.5$, these terms cannot
contribute significantly to the deceleration parameter. 

In the first term in Eq. (17), the quantities $\rho$ and $2\nabla_{\alpha}\phi\nabla_{\beta}\phi u^{\alpha}u^{\beta}$
are always positive (the second term is the square of $\sqrt{2}\nabla\phi_{\alpha}u^{\alpha}$).
This means that the potential term has to be larger than the sum of
these two terms to get negative deceleration. Thus, a large negative
deceleration must come primarily from the potential.

We note that this result differs from that obtained by Martineau and
Brandenberger in Ref. {[}50], who found that the backreaction of superhorizon
perturbations $\mathit{could}$ drive cosmic acceleration via a mechanism
not involving the potential. A possible reason for the difference
is the fact that different measures of cosmic acceleration are used
in the two different analyses. The authors of Ref. {[}50] use a measure
that is based on averages over a spatial slice at a given instant
of time (which is inherently gauge dependent). We use a different
measure which is essentially an average over the past light cone of
the observer, and is gauge independent. Moreover, our measure corresponds
more closely to the actual deceleration parameter that has been measured.

Finally, we note that the specific choices of angle averaging prescriptions
in the definitions (8) of the Hubble parameter and deceleration parameter
are not unique. However, as was argued in Ref. {[}31], the change
that results from adopting other definitions is negligible. For example,
one could consider the alternative definition 

\begin{equation}
q_{0}\equiv1-2H_{0}\left\langle B\right\rangle \label{eq:18}\end{equation}
of the deceleration parameter. In Ref. {[}31] it was shown that this
alters the final result (17) in three ways: (i) Changing the numerical
coefficients of the shear squared and vorticity squared terms by an
amount of order unity, which does not affect our conclusions; (ii)
The addition of new terms that are comparable to the shear squared
and vorticity squared terms; and (iii) The addition of new terms that
are suppressed compared to the shear squared and vorticity squared
terms by one or more powers of the dimensionless ratio (non-isotropic
part of $\nabla_{\alpha}u_{\beta}$)/(isotropic part of $\nabla_{\alpha}u_{\beta}$).
This dimensionless ratio is constrained observationally to be small
compared to unity, since peculiar velocities on Hubble scales today
are small. The same arguments continue to apply in the present context,
since the scalar field dependent terms in (17) are unchanged by the
change in definition of $q_{0}.$

\section{Conclusion}

~~The backreaction of perturbations is sometimes considered to be
a candidate for explaining cosmic acceleration {[}53]. Many techniques
have been developed to explore the effects of backreaction. In this
paper, we computed the deceleration parameter measured by comoving
observers in a hypothetical universe with all perturbation modes which
are subhorizon today set to zero at early times. We considered a universe
containing cold dark matter and a minimally coupled scalar field.
We showed that one can obtain a large negative value of the deceleration
parameter in this context only if the deceleration is primarily produced
by the scalar field's potential.

\end{document}